\documentclass[a4paper,12pt]{article}

\usepackage[pdftex]{graphicx}
\usepackage[colorlinks]{hyperref}

\providecommand{\keywords}[1]
{
  \small
  \textbf{Keywords:} #1
}

\usepackage{amsmath,amssymb,amsthm}
\usepackage{bm} 
\usepackage{mathtools}
\usepackage{booktabs}
\usepackage{cite}
\usepackage{comment}

\theoremstyle{definition}
\newtheorem{example}{Example}
\AtBeginEnvironment{example}{%
	\pushQED{\qed}%
}
\AtEndEnvironment{example}{\popQED\endexample}

\makeatletter
    
    \@addtoreset{equation}{section}
\makeatother

\title{\Large\textbf{Interrelation among Solvable Potentials and Extensions of SWKB Quantization Condition}}
\author{\large Yuta Nasuda\thanks{
	E-mail: \texttt{y.nasuda.phys@gmail.com}, ORCiD: \texttt{0000-0002-0149-8483}}
	\\[1ex] 
	{\normalsize Department of Natural Sciences, National Institute of Technology, Gunma College,} \\
	{\normalsize 580 Toriba, Maebashi, Gunma 371-8530, Japan}
	}
\date{\footnotesize (Dated: \today)}

\allowdisplaybreaks[4]

\begin{document}

\maketitle

\begin{abstract}
The exactly solvable Schr\"{o}dinger equations with the conventional shape-invariant potentials are known to be related with each other through point cannonical transformations.
In this paper, we extend the idea to integral formulae called the SWKB integrals.
By virtue of this, we derive extended forms of the SWKB quantization condition for certain classes of Natanzon potentials.
We further demonstrate that the same idea can also be applied to obtain an exact quantization rule for a subclass of quantum systems with position-dependent effective masses, provided their solutions involve the classical orthogonal polynomials.
Based on the findings, we conjecture about the implication of the exactness of the SWKB formula in relation to the classical orthogonal polynomials.
\end{abstract}

\keywords{
Schr\"{o}dinger equation, classical orthogonal polynomials, quantization condition
}

\section{Introduction}
\label{sec:Intro}
In mathematical sciences, orthogonal polynomials play a fundamental role. 
They appear in the exact solutions of various models, which enables qualitative analyses and reveals underlying mathematical structure behind phenomena. 
The classical orthogonal polynomials (COPs)~\cite{alma9910116860608801} are particularly valued for their simplicity and tractability.

The significance of the classical orthogonal polynomials is evident especially in quantum mechanics, where one can construct a potential in such a way that the eigenfunctions of the Hamiltonian with the potential are expressed in terms of a given orthogonal polynomials. 
The most familiar example is the 1-dim. harmonic oscillator, whose eigenfunctions involve the Hermite polynomials (H). 
Other typical examples are the radial oscillator with the Laguerre polynomials (L), and the P\"{o}schl--Teller potential with the Jacobi polynomials (J).

Researchers have been exploring novel exactly solvable potentials in quantum mechanics. 
Unfortunately, a no-go theorem called Bochner's theorem~\cite{routh1884some,Bochner:1929aa} tells us that there are no essentially new classes of exactly solvable potentials in terms of ``ordinary'' orthogonal polynomials.
Some considered other special functions such as the Bessel functions~\cite{Sasaki_2016,doi:10.1142/S0219887825400304}, while others loosen the antecedent of the no-go theorem.
The latter led to the developments of the Krein--Adler transformation~\cite{krein1957continuous,adler1994modification}, the multi-indexed orthogonal polynomials~\cite{sasaki2010exceptional,ODAKE2011164} (including the exceptional orthogonal polynomials~\cite{GOMEZULLATE2009352,GOMEZULLATE2010987,quesne2008exceptional,ODAKE2009414,ODAKE2010173}).  
They overcame Bochner's theorem in the sense that those orthogonal polynomials lack several degrees. 
Alternative approaches involve modifications of the governing equation.
Second-order linear differential equations with non-constant coefficients (See below.) and difference equations (See, \textit{e.g.}, Ref.~\cite{Odake_2011}.) have been considered.

Yet another approach was proposed by Natanzon.
A class of exactly solvable potentials named after him is a class of exactly solvable potentials that has finite number of discrete eigenvalues and the corresponding eigenfunction is expressed by the (confluent) hypergeometric function(s)~\cite{natanzon1971study,natanzon1979general,ginocchio1984class,cooper1987relationship}. 
One feature of the eigenvalues and eigenfunctions of this class is that they are typically obtained in an implicit rather than explicit form.
It turned out that the Schr\"{o}dinger equations for this class of potentials are connected to those for the conventional shape-invariant potentials by changing variables, \textit{i.e.}, point canonical transformations.

The transformations of the conventional shape-invariant potentials yield another class of exactly solvable quantum-mechanical systems~\cite{quesne-SIGMA2007}. 
They are quantum-mechanical systems with position-dependent effective masses, where the coefficient of the second-derivative term has position dependency rather than being constant~\cite{PhysRev.152.683,PhysRevB.27.7547}. 
Such systems solved by the classical orthogonal polynomials were first systematically constructed by Bagchi \textit{et al.}~\cite{Bagchi_2005}
We note here that there is a quantum-mechanical system with position-dependent effective mass solved by the classical orthogonal polynomials which cannot be obtained from the conventional shape-invariant ones by point canonical transformations~\cite{Jafarov-2021,Jafarov:2022aa}.

The exact solvability of the Schr\"{o}dinger equation is often discussed in the context of supersymmetric quantum mechanics~\cite{witten1981dynamical,witten1982constraints,cooper1995supersymmetry}. 
For the potentials solved by the classical orthogonal polynomials such as the 1-dim. harmonic oscillator (H), the radial oscillator (L) and P\"{o}schl--Teller potential (J), Gendenstein revealed underlying symmetry known as shape invariance~\cite{gendenshtein1983derivation}. 
By respecting this symmetry, those potentials are often referred to as the conventional shape-invariant potentials. 
(One can find the list of the potentials of this class in, \textit{e.g.}, Refs.~\cite{cooper1995supersymmetry,gangopadhyaya2017supersymmetric}.)

In 1985, Comtet and his co-authors proposed an exact quantization condition inspired by the semi-classical approximation, for the conventional shape-invariant potentials~\cite{comtet1985exactness}. 
The condition is often referred to as the SWKB quantization condition. 
(For a recent review, see Ref.~\cite{doi:10.1142/S0218271825300010}.)
After several case studies~\cite{dutt1986exactness,khare1989shape,delaney1990susy,bougie2018supersymmetric,Nasuda:2020aqf,10.1007/978-981-19-4751-3_29,NASUDA2023116087}, it has been proved that this condition is exact if and only if the potential is a conventional shape-invariant potential. 
For other potentials, the condition gives a good approximation of the energy eigenvalues. 
Bhalla \textit{et al.}~\cite{bhalla1996exactness,bhalla1997quantum} explained the exactness (and the non-exactness) of the SWKB quantization condition by comparing it with the exact quantization condition derived in the quantum Hamilton--Jacobi theory~\cite{leacock1983hamilton_action,leacock1983hamilton,gozzi1986nodal}. 
The (non-)exactness is accounted for by the poles and branch cuts of the quantum momentum function, which is proportional to the logarithmic derivative of the wavefunction. 
This implies that the SWKB condition is not only the quantization of a certain quantity (which we call the SWKB integral), ora tool for spectral computation, but also a probe into exploring the analytic structure of the wavefunctions, and therefore the special functions.

This naturally leads to the following question.
Is there any way of extending the SWKB quantization condition so that it reproduces exact bound-state spectra for other potentials than conventional shape-invariant ones? 
In the present paper, we propose such extensions of the condition equation that it will be exact for certain classes of exactly solvable potentials. 
By incorporating point canonical transformations, we derive exact quantization conditions.

This Paper is organized as follows.
In the subsequent section \ref{sec:SI-SWKB} is devoted for a quick review on the concept of the shape invariance and the SWKB quantization condition.
Then in Sect. \ref{sec:cSI}, three explicit and elementary exapmles of the potenstial with shape invariance are provided.
Interrelations among the conventional shape-invariant potentials are also recapitulated here.
The interconnections in terms of the SWKB integrals are one of the main results of this work.
Sect. \ref{sec:Natanzon} is the review on certain classes of the Natanzon potentials.
Sects. \ref{sec:ext-SWKB} and \ref{sec:dSWKB} are the other highlights of this paper.
The former studies exact quantization rules for the Natanzon potentials, while the latter is for  classes with position-dependent effective masses.
We summarize our analyses in Sect. \ref{sec:Conclusion} to arrive at a conjecture about the implication of the exactness of the SWKB formula in relation to the classical orthogonal polynomials.

\section{Shape Invariance and SWKB Quantization Condition}
\label{sec:SI-SWKB}
In this paper, we deal with the factorized Hamiltonian:
\begin{equation}
\mathcal{H} = -\frac{d^2}{dx^2} + V(x)
= \left( -\frac{d}{dx} + W(x) \right) \left( \frac{d}{dx} + W(x) \right) ~,~~~
V(x) \equiv W(x)^2 - \frac{dW(x)}{dx} ~,
\end{equation}
where $V(x)$ is the potential and $W(x)$ is often referred to as the superpotentioal.
We forcus on the case where the Hamiltonians are bounded from below so that they have infinitely or finitely many discrete eigenvalues:
\begin{equation}
\mathcal{H}\psi_n(x) = \mathcal{E}_n\psi_n(x) ~,~~~
n = 0,1,2,\ldots ~.
\label{eq:SE}
\end{equation}
The factorized Hamiltonian corresponds to the vanishing ground-state energy, $\mathcal{E}_0 = 0$.
The numbering of the eigenvalues is monotonously increasing, $0 = \mathcal{E}_0 < \mathcal{E}_1 < \mathcal{E}_2 < \cdots$.

A sufficient condition for the exact solvability of the eigenvalue problem \eqref{eq:SE} called shape invariance is well discussed in the context of superpymmetric quantum mechanics.
The potential is said to be shape invariant, if
\begin{equation}
W(x;\bm{a})^2 + \frac{dW(x;\bm{a})}{dx} = W(x;f(\bm{a}))^2 - \frac{dW(x;f(\bm{a}))}{dx} + \epsilon (\bm{a}) ~,
\label{eq:SI}
\end{equation}
in which $\bm{a} = \{ a_1, a_2, \ldots \}$ is the set of parameters in the potential, and the parameter dependency of nthe superpotential is denoted explicitly, $W(x) \equiv W(x;\bm{a})$.
$f$ is some function, \textit{e.g.}, $f(a) = a+1$, and $\epsilon (\bm{a})$ is a constant.
Explicit examples of potentials with this symmetry is presented later in Sec. \ref{sec:cSI}.
One can also find the list in the textbooks such as Refs. \cite{cooper2001supersymmetry,gangopadhyaya2017supersymmetric}.
Other exaples are found in Refs. \cite{quesne2008exceptional,ODAKE2009414,ODAKE2010173,sasaki2010exceptional,ODAKE2011164}.

Comtet, Bandrauk and Campbell proposed a WKB-like quantization condition, which reads
\begin{equation}
\int_a^b \sqrt{\mathcal{E}_n - W(x)^2} \, dx = n\pi ~,~~~
n=0,1,2,\ldots ~,
\end{equation}
with $a$ and $b$ being the two solutions of the equation $\mathcal{E}_n = W(x)^2$.
This condition is often referred to as the SWKB quantization condition, and the integral on the l.h.s. is sometimes called the SWKB integral.
A notable feature of this condition is that it reproduces exact bound-state spectra for a class of exactly solvable potentials.

\section{Generating Novel Solvable Potentials from Conventional Ones}
\label{sec:3}
\subsection{Conventional shape-invariant potentials}
\label{sec:cSI}
An elementary class of exactly solvable potentials is the conventional shape-invariant potentials, which respect the condition equation \eqref{eq:SI}, and the term ``conventional'' reflects the fact that these systems were already known in the 1950's.
Here are the three explicit examples of the conventional shape-invariant potentials:
\begin{align}
\text{H:}\quad &\left[ -\frac{d^2}{dz^2} + \omega^2z^2 - \omega \right] \phi_n(z) = 2n\omega\phi_n(z) ~,~~~ 
-\infty < z < \infty ~, \nonumber \\
&\qquad \phi_n(z) = \mathrm{e}^{-\frac{\omega z^2}{2}}H_n(\sqrt{\omega}z) ~,~~~
\mathcal{E}_n = 2n\omega ~,
\label{eq:SE-H} \\
\text{L:}\quad &\left[ -\frac{d^2}{dz^2} + \frac{g(g-1)}{z^2} + \omega^2z^2 - \omega (2g + 1) \right] \phi_n(z) = 4n\omega\phi_n(z) ~,~~
0 < z < \infty ~, \nonumber \\
&\qquad 
\phi_n(z) = \mathrm{e}^{-\frac{\omega z^2}{2}}z^g L_n^{(g-\frac{1}{2})}(\omega z^2) ~,~~~
\mathcal{E}_n = 4n\omega ~,
\label{eq:SE-L} \\
\text{J:}\quad &\left[ -\frac{d^2}{dz^2} + \frac{g(g-1)}{\sin^2 z} + \frac{h(h-1)}{\cos^2 z} - (g + h)^2 \right] \phi_n(z) = 4n(n + g + h)\phi_n(z) ~,~~
0 < z < \frac{\pi}{2} ~, \nonumber \\
&\qquad
\phi_n(z) = (\sin z)^g (\cos z)^h P_n^{(g-\frac{1}{2},h-\frac{1}{2})}(\cos 2z) ~,~~
\mathcal{E}_n = 4n(n+g+h) ~.
\label{eq:SE-J} 
\end{align}
In the above potentials, $\omega$, $g$, $h$ are model parameters with $\omega > 0$ and $g,h > 1/2$.

The three are typical in the sense that all conventional shape-invariant potentials are generated by chages of variables from the three. 
(See Tab. \ref{tab:Interrel_cSI-SE} as well as Example \ref{ex:R2C-SE}.)
In the other way around, all conventional shape-invariant potentials are reduced into the three.
This is accounted for by the fact that their Schr\"{o}dinger equations are solved by classical orthogonal polynomials.

\begin{example}[Radial oscillator to Coulomb potential]
\label{ex:R2C-SE}
The Schr\"{o}dinger equation for the radial oscillator \eqref{eq:SE-L} transforms into one for the Coulomb potential, under the change of variables:
\begin{equation}
z \mapsto x = z^2
\qquad\text{\it i.e.}\quad
z = \sqrt{x} ~.
\end{equation}
In the meanwhile, we write the wavefunction $\phi_n(z) \mapsto \varphi_n(x) = \sqrt{2z}\,\phi_n(z)$ and also the paramters
\begin{equation}
\omega = \frac{e^2}{(\tilde{g}+n)} ~,~~~ g = 2\tilde{g} - \frac{1}{2} ~.
\end{equation}
The resulting differential equation is
\begin{equation}
\left[ -\frac{d^2}{dx^2} + \frac{\tilde{g}(\tilde{g}-1)}{x^2} - \frac{e^2}{x} + \frac{e^4}{4\tilde{g}^2} \right]\psi_n(x) = \left( \frac{e^4}{4\tilde{g}^2} - \frac{e^4}{4(\tilde{g}+n)^2} \right) \psi_n(x) ~,
\end{equation}
with
\begin{equation}
\psi_n(x) = \mathrm{e}^{-\frac{e^2x}{2(\tilde{g}+n)}}x^gL_n^{(2\tilde{g}-1)}\left( \frac{e^2}{\tilde{g}+n}x \right) ~.
\end{equation}
\end{example}

\begin{table}[t]
\centering
\caption{Interrelation among the Schr\"{o}dinger equations of conventional shape-invariant potentials.}
\label{tab:Interrel_cSI-SE}
\smallskip
\scalebox{0.7}{
	\begin{tabular}{cccccl}
	\toprule
	Class & Superpotential & Energy &~~& \begin{tabular}{c}
			Change of variables~~ \\
			$x \mapsto z=z(x)$~~
		\end{tabular} & Relation among parameters \\
	\midrule
	H & \begin{tabular}{c}
			1-dim. HO \\
			$\omega z$ 
		\end{tabular} & $2n\omega$ && --- & --- \\
	\midrule
	L & \begin{tabular}{c}
			Radial osc. \\
			$\omega z - \dfrac{ g}{z}$ 
		\end{tabular} & $4n\omega$ && --- & --- \\
	  & \begin{tabular}{c}
	  		Coulomb pot. \\
			$\dfrac{e^2}{2\tilde{g}} - \dfrac{\tilde{g}}{x}$ 
		\end{tabular} & $\dfrac{e^4}{4\tilde{g}^2} - \dfrac{e^4}{4(\tilde{g}+n)^2}$ && $\sqrt{x}$ & $\omega = \dfrac{e^2}{(g+n)}$, $g = 2\tilde{g}-\dfrac{1}{2}$ \\
	  & \begin{tabular}{c}
	  		Morse pot. \\
			$\mu\mathrm{e}^x - h$ 
		\end{tabular} & $2nh-n^2$ && $\mathrm{e}^{x/2}$ & $\omega = 2\mu$, $g = 2(h-n)+\dfrac{1}{2}$ \\
	\midrule
	J & \begin{tabular}{c}
			P\"{o}schl--Teller pot. \\
			$- g\cot z + h\tan z$ 
		 \end{tabular} & $4n(n+g+h)$ && --- & --- \\
	   & \begin{tabular}{c}
			Hyperbolic symmetric top II \\
			$\dfrac{\mu}{\cosh x} + \tilde{h}\tanh x$ 
		 \end{tabular} & $2n\tilde{h}-n^2$ && $\dfrac{1}{2}\arccos (\mathrm{i}\sinh x)$ & $g = -\tilde{h}-\mu\mathrm{i}$, $h = -\tilde{h}+\mu\mathrm{i}$ \\
	   & \begin{tabular}{c}
			Rosen--Morse pot. \\
			$\dfrac{\mu}{\tilde{h}} + \tilde{h}\tanh x$ 
		 \end{tabular} & $2n\tilde{h}-n^2 + \dfrac{\mu^2}{\tilde{h}^2} - \dfrac{\mu^2}{(\tilde{h}-n)^2}$ && $\dfrac{1}{2}\arccos (\tanh x)$ & \begin{tabular}{l} 
		 	$g = \tilde{h}-n+\dfrac{\mu}{\tilde{h}-n}+\dfrac{1}{2}$, \\
			$h = \tilde{h}-n-\dfrac{\mu}{\tilde{h}-n}+\dfrac{1}{2}$
		 \end{tabular} \\
	   & \begin{tabular}{c}
			Eckart pot. \\
			$\dfrac{\mu}{\tilde{g}} - \tilde{g}\coth x$ 
		 \end{tabular} & $2n\tilde{g}-n^2 + \dfrac{\mu^2}{\tilde{g}^2} - \dfrac{\mu^2}{(\tilde{g}+n)^2}$ && $\dfrac{1}{2}\arccos (\coth x)$ & \begin{tabular}{l} 
		 	$g = -\tilde{g}-n+\dfrac{\mu}{\tilde{g}+n}+\dfrac{1}{2}$, \\
			$h = -\tilde{g}-n-\dfrac{\mu}{\tilde{g}+n}+\dfrac{1}{2}$
		 \end{tabular} \\
	   & \begin{tabular}{c}
			Hyperbolic P\"{o}schl--Teller pot. \\
			$- g\coth x + \tilde{h}\tanh x$ 
		 \end{tabular} & $4n(\tilde{h}-g-n)$ && $\arcsin (-\mathrm{i}\sinh x)$ & $h = -\tilde{h}$ \\
	\bottomrule
	\end{tabular}
}
\end{table}

\subsubsection{In terms of SWKB quantization conditions}
\label{sec:Interrel-SWKB}
There are also similar interrelations among the SWKB integrals for conventional shape-invariant potentials.
All the SWKB integrals for conventional shape-invariant potentials are generated by change of variables from those for the three typical potentials:
\begin{align}
\text{H:}\qquad &\int_{\alpha_-}^{\alpha_+} \sqrt{2n\omega - \omega^2z^2} \, dz = n\pi ~,~~~
\alpha_{\pm} = \pm\sqrt{\frac{2n}{\omega}} ~, 
\label{eq:SWKB-H} \\
\text{L:}\qquad &\int_{\alpha_-}^{\alpha_+} \sqrt{4n\omega - \left( \omega z -\frac{g}{z} \right)^2} \, dz = n\pi ~,~~~
\alpha_{\pm} = \frac{\sqrt{n+g} \pm \sqrt{n}}{\sqrt{\omega}} ~, 
\label{eq:SWKB-L} \\
\text{J:}\qquad &\int_{\alpha_-}^{\alpha_+} \sqrt{4n(n+g+h) - \left( -g\cot z + h\tan z \right)^2} \, dz = n\pi ~, \nonumber \\
&\hspace{.3\linewidth} \alpha_{\pm} = \arctan \frac{\sqrt{(n+g)(n+h)} \pm \sqrt{n(n+g+h)}}{h} ~,
\label{eq:SWKB-J}
\end{align}
with $n$ being a non-negative integer.
(See Tab. \ref{tab:Interrel_cSI-SWKB} as well as Example \ref{ex:R2C-SWKB}.)
Note that these three formulae \eqref{eq:SWKB-H}--\eqref{eq:SWKB-J} are demonstrated by analytical direct calculation.
In the other way around, any SWKB integrals for conventional shape-invariant potentials are mapped into the integral for either (H), (L) or (J) by the same chages of variables as in the case of Schr\"{o}dinger equations.
We point out here that this gives another proof of the exactness of SWKB quantization condition.

\begin{table}[t]
\centering
\caption{Interrelations among the SWKB integrals for conventional shape-invariant potentials.}\label{tab:Interrel_cSI-SWKB}
\smallskip
\scalebox{0.7}{
	\begin{tabular}{cccccl}
	\toprule
	Class & Superpotential & Energy &~~& \begin{tabular}{c}
			Change of variables~~ \\
			$x \mapsto z=z(x)$~~
		\end{tabular} & Relation among parameters \\
	\midrule
	H & \begin{tabular}{c}
			1-dim. HO \\
			$\omega z$ 
		\end{tabular} & $2n\omega$ && --- & --- \\
	\midrule
	L & \begin{tabular}{c}
			Radial osc. \\
			$\omega z - \dfrac{ g}{z}$ 
		\end{tabular} & $4n\omega$ && --- & --- \\
	  & \begin{tabular}{c}
	  		Coulomb pot. \\
			$\dfrac{e^2}{2\tilde{g}} - \dfrac{\tilde{g}}{x}$ 
		\end{tabular} & $\dfrac{e^4}{4\tilde{g}^2} - \dfrac{e^4}{4(\tilde{g}+n)^2}$ && $\sqrt{x}$ & $\omega = \dfrac{e^2}{(g+n)}$, $g = 2\tilde{g}$ \\
	  & \begin{tabular}{c}
	  		Morse pot. \\
			$\mu\mathrm{e}^x - h$ 
		\end{tabular} & $2nh-n^2$ && $\mathrm{e}^{x/2}$ & $\omega = 2\mu$, $g = 2(h-n)$ \\
	\midrule
	J & \begin{tabular}{c}
			P\"{o}schl--Teller pot. \\
			$-g\cot z + h\tan z$ 
		 \end{tabular} & $4n(n+g+h)$ && --- & --- \\
	   & \begin{tabular}{c}
			Hyperbolic symmetric top II \\
			$\dfrac{\mu}{\cosh x} + \tilde{h}\tanh x$ 
		 \end{tabular} & $2n\tilde{h}-n^2$ && $\dfrac{1}{2}\arccos (\mathrm{i}\sinh x)$ & $g = -\tilde{h}-\mu\mathrm{i}$, $h = -\tilde{h}+\mu\mathrm{i}$ \\
	   & \begin{tabular}{c}
			Rosen--Morse pot. \\
			$\dfrac{\mu}{\tilde{h}} + \tilde{h}\tanh x$ 
		 \end{tabular} & $2n\tilde{h}-n^2 + \dfrac{\mu^2}{\tilde{h}^2} - \dfrac{\mu^2}{(\tilde{h}-n)^2}$ && $\dfrac{1}{2}\arccos (\tanh x)$ & \begin{tabular}{l} 
		 	$g = \tilde{h}-n+\dfrac{\mu}{\tilde{h}-n}$, \\
			$h = \tilde{h}-n-\dfrac{\mu}{\tilde{h}-n}$
		 \end{tabular} \\
	   & \begin{tabular}{c}
			Eckart pot. \\
			$\dfrac{\mu}{\tilde{g}} - \tilde{g}\coth x$ 
		 \end{tabular} & $2n\tilde{g}-n^2 + \dfrac{\mu^2}{\tilde{g}^2} - \dfrac{\mu^2}{(\tilde{g}+n)^2}$ && $\dfrac{1}{2}\arccos (\coth x)$ & \begin{tabular}{l} 
		 	$g = -\tilde{g}-n+\dfrac{\mu}{\tilde{g}+n}$, \\
			$h = -\tilde{g}-n-\dfrac{\mu}{\tilde{g}+n}$
		 \end{tabular} \\
	   & \begin{tabular}{c}
			Hyperbolic P\"{o}schl--Teller pot. \\
			$-g\coth x + \tilde{h}\tanh x$ 
		 \end{tabular} & $4n(\tilde{h}-g-n)$ && $\arcsin (-\mathrm{i}\sinh x)$ & $h = -\tilde{h}$ \\
	\bottomrule
	\end{tabular}
}
\end{table}

\begin{example}[Radial oscillator to Coulomb potential]
\label{ex:R2C-SWKB}
The SWKB integral for the radial oscillator \eqref{eq:SWKB-L} transforms into that for the Coulomb potential, under the same change of variables as in the case of Example \ref{ex:R2C-SE}:
\begin{equation}
z \mapsto x = z^2
\qquad\text{\it i.e.}\quad
z = \sqrt{x} ~.
\end{equation}
The replacements of paramters are slitely different,
\begin{equation}
\omega = \frac{e^2}{(\tilde{g}+n)} ~,~~~ g = 2\tilde{g} ~.
\end{equation}
The resulting integral is
\begin{equation}
\int_{a_-}^{a_+} \sqrt{-\frac{e^4}{4(\tilde{g}+n)^2} + \frac{e^2}{x} - \frac{\tilde{g}^2}{x^2}} \, dx
= \int_{a_-}^{a_+} \sqrt{\frac{e^4}{4\tilde{g}^2} - \frac{e^4}{4(\tilde{g}+n)^2} - \left( \frac{e^2}{2\tilde{g}} - \frac{\tilde{g}}{x} \right)^2} \, dx ~.
\end{equation}

\end{example}

Since the same changes of variables as in the case of Schr\"{o}dinger equations realize the mappings of conventional shape-invariant SWKB integrals, one can guess that this mapping is also accounted for by the solvability by the classical orthogonal polynomials.
In what follows, we are to extend the quantization condition to be exact for any potentials involve the classical orthogonal polynomials.
For this purpose, we first consider a general class of exactly solvable potentials where the main part of the wavefunction is expressed by a single classical orthogonal polynomial, which are known as the Natanzon potentials.

\subsection{Natanzon potentials}
\label{sec:Natanzon}
This subsection is devoted to the review of a construction of the Natanzon potentials in relation to conventional shape-invariant potentials by changes of variables.

\subsubsection{Laguerre}
One class of Natanzon potentials is constructed from the radial oscillator by the following procedure.
We start from the Schr\"{o}dinger equation for the radial oscillator \eqref{eq:SE-L}, symbolically we write
\begin{equation}
\left[ -\frac{d^2}{dz^2} + U(z) \right] \phi_n(z) = \varepsilon_n\phi_n(z) ~.
\end{equation}
The coordinate transformation:
\begin{equation}
z\mapsto x=x(z)
\qquad\text{with}\quad
\frac{dx}{dz} = f(x) ~,
\label{eq:dx/dz=f}
\end{equation}
and the rescaling of the wave function: $\phi\mapsto\psi = \sqrt{f}\,\phi$, yeild
\begin{equation}
\left[ -\frac{d^2}{dx^2} - \frac{f'^2}{4f^2} + \frac{f''}{2f} + \frac{U - \varepsilon_n}{f^2} \right] \psi(x) = 0 ~. 
\end{equation}
By adding a trem $E\psi(x)$ on the both sides of the equation, we obtain
\begin{equation}
\left[ -\frac{d^2}{dx^2} + V(x) \right] \psi(x) 
= \left[ -\frac{d^2}{dx^2} - \frac{f'^2}{4f^2} + \frac{f''}{2f} + \frac{U - \varepsilon_n + Ef^2}{f^2} \right] \psi(x)
= E\psi(x) ~.
\end{equation}
In order for this equation to be the eigenvalue equation for the Hamiltonian $-d^2/dx^2 + V(x)$, it is necessary that the numerator of the third term in the potential, $g\coloneqq U - \varepsilon_n + Ef^2$, is independent of $n$.
This condition implies that $f^2$ and $g$ have the same functional form as $U$, that is
\begin{align}
f(x) &= f(z(x)) = \sqrt{\frac{A}{z^2} + Bz^2 + C} ~, 
\label{eq:NatanzonL-f} \\
g(z) &= U(z) - \varepsilon_n + Ef^2 = \frac{D}{z^2} + Fz^2 + G ~.
\end{align}
For this $f(x)$ \eqref{eq:NatanzonL-f}, the solution of the differential equation \eqref{eq:dx/dz=f} is
\begin{multline}
x =
z \left[ \frac{\sqrt{A + Cz^2 + Bz^4}}{2} -\frac{C}{4\sqrt{B}} \ln \left(-2 \sqrt{B} \sqrt{A + Cz^2 + Bz^4} + 2Bz^2 + C \right) \right. \\
\left. - \sqrt{A} \tanh^{-1} \left( \frac{\sqrt{A + Cz^2 + Bz^4} - \sqrt{B}z^2}{\sqrt{A}}\right)\right]
\label{eq:NatanzonL-x}
\end{multline}
Note that, in general, Eq. \eqref{eq:NatanzonL-x} cannot be solved explicitly for $z$ as a function of $x$, meaning that we only know $V(x)$ implicitly.

\subsubsection{Jacobi}
Another class of Natanzon potentials is made from the P\"{o}schl--Teller potential.
In this case, we start with the P\"{o}schl--Teller potential \eqref{eq:SE-L}.
The construction procedure is almost identical to the Laguerre case, but with
\begin{equation}
f(x) = f(z(x)) = \sqrt{\frac{A}{\sin^2 x} + \frac{B}{\cos^2 z} + C} ~, 
\end{equation}
meaning
\begin{multline}
x = \frac{1}{2} 
\left[ \sqrt{A} \coth ^{-1}\left( \frac{2 \sqrt{A} \sqrt{A+ \left(A + B\sec^2 z + C \right)\tan^2 z}}{2A + (A+B+C)\tan^2 z}\right) \right. \\
-\sqrt{B} \coth ^{-1}\left( \frac{2 \sqrt{B} \sqrt{A+ \left(A + B\sec^2 z + C \right)\tan^2 z}}{A+B+C + 2B \tan^2 z} \right) \\
\left. +\sqrt{C} \cot ^{-1}\left( \frac{2\sqrt{C}\cos^2 z \sqrt{A+ \left(A + B\sec^2 z + C \right)\tan^2 z}}{-A + B + C\cos 2z} \right) \right] ~,
\end{multline}
and
\begin{equation}
g(z) = U(z) - \varepsilon_n + Ef^2 = \frac{D}{\sin^2 z} + \frac{F}{\cos^2 z} + G ~.
\end{equation}

Before closing this section, we note that naive application of the SWKB quantization condition does not reproduce exact bound-state spectra for Natanzon potentials.
This has already been discussed in the literature \cite{khare1989shape}.

\section{Extension of SWKB Quantization Condition}
\label{sec:ext-SWKB}
As we have seen in Sec. \ref{sec:Interrel-SWKB}, under certain changes of variables and replacements of parameters, the integrands in Eqs. \eqref{eq:SWKB-H}--\eqref{eq:SWKB-J} are organized in the following form:
\[
\sqrt{\left[ \left\{ \begin{array}{c}
	\text{$n$-dependent} \\
	\text{and} \\
	\text{$x$-independent}
\end{array} \right\} \text{term(s)} \right] - \left[ \left\{ \begin{array}{c}
	\text{$n$-independent} \\
	\text{and} \\
	\text{$x$-dependent}
\end{array} \right\} \text{term(s)} \right]^2} ~.
\]
A remarkable feature in the case of conventional shape-invariant systems is that ``$n$-dependent and $x$-independent term(s)'' are equal to the energy eigenvalues and ``$n$-independent and $x$-dependent term(s)'' coincide with the superpotential.
On the other hand, the fact that the SWKB condition does not reproduce exact bound-state spectra for Natanzon potentials imples that this correspondence is not the case.

However, these is still a possibility that there is another form of integral formula (or quantization condition) for Natanzon potentials, which is derived from the SWKB formulae for (L) and (J).

\subsection{Laguerre}

Since Natanzon potentials are, genarally, only expressed inplicitly in terms of $z$ instead of $x$, let us first write the quantization condition in $z$:
\begin{equation}
\int_{\alpha}^{\beta} \sqrt{4n\sqrt{F - BE_n} - \left( \sqrt{F - BE_n}z - \frac{1 + \sqrt{1 + 4D - AE_n}}{2z} \right)^2} \,dz = n\pi ~.
\end{equation}

Now let us rewrite this equation in $x$.
Since $dz = f(x)\,dx$ and $-\dfrac{d}{dz}\ln\phi_0(z) = -\dfrac{1}{f(x)}\dfrac{d}{dx}\ln\dfrac{\psi_0(x)}{\sqrt{f(x)}}$, formally we write
\begin{equation}
\int_a^b \sqrt{\frac{4n\sqrt{F - BE_n}}{f(x)} - \left( W(x) + \frac{f'(x)}{2f(x)} \right)^2} \,dx 
= n\pi ~.
\end{equation}

\subsection{Jacobi}
Similary, for Jacobi case,
\begin{multline}
\int_{\alpha}^{\beta} \Biggl\{ 2n\left( 2n + 2+\sqrt{1 - 4AE_n + 4D} + \sqrt{1 - 4BE_n + 4F} \right) \\ 
- \left[ -\frac{1}{2}\left( 1 + \sqrt{1 - 4AE_n + 4D} \right) \cot z + \frac{1}{2}\left( 1 + \sqrt{1 - 4BE_n + 4F} \right) \tan z \right]^2 \Biggr\}^{\frac{1}{2}} \,dz = n\pi ~,
\end{multline}
or
\begin{equation}
\int_a^b \sqrt{\frac{2n\left( 2n + 2+\sqrt{1 - 4AE_n + 4D} + \sqrt{1 - 4BE_n + 4F} \right)}{f(x)} - \left( W(x) + \frac{f'(x)}{2f(x)} \right)^2} \,dx 
= n\pi ~.
\end{equation}

\section{SWKB Quantization Condition with Position-dependent Effective Masses}
\label{sec:dSWKB}
In constructing Natanzon potentials, we have required that the coefficient of the second-derivative term is constant under change of variables.
However, it is also worth examining the case where the coefficient depends on $x$, which corresponds to a quantum mechanical system with position-dependent masses.

\subsection{Deformed shape-invariant potentials}
In this subsection, we revisit a deformation of SUSY QM with position-dependent effective masses \cite{Bagchi_2005,quesne-SIGMA2007,sym12111853}.
We define the deformed Hamiltonian $\mathcal{H}_{\rm d}$ in a factorized form:
\begin{align}
\mathcal{H}_{\rm d} &= \left( -\frac{1}{\sqrt{2m_0}}\sqrt{\eta(x)}\frac{d}{dx}\sqrt{\eta(x)} + W(x) \right) \left( \frac{1}{\sqrt{2m_0}}\sqrt{\eta(x)}\frac{d}{dx}\sqrt{\eta(x)} + W(x) \right) ~, \\
&= -\frac{1}{2m_0}\left(\sqrt{\eta(x)}\frac{d}{dx}\sqrt{\eta(x)}\right)^2 + V_{\mathrm{eff}}(x) ~,~~~
V_{\mathrm{eff}}(x) = W(x)^2 - \frac{\eta(x)}{\sqrt{2m_0}}\frac{dW(x)}{dx} ~.
\end{align}
Here, $\eta(x)$ describes the deformation of the system; the limit $\eta(x) \to 1$ corresponds to undeformed, ordinary quantum-mechanical Hamiltonian.
This deformation of the Hamiltonian can be interpreted not only as the system with position-dependent effective mass, but also as the mapping of Schr\"{o}dinger equation to a different Sturm--Liouville eigenvalue problem.
The eigenvalue equation for $\mathcal{H}_{\rm d}$ is
\begin{equation}
\mathcal{H}_{\rm d}\psi_n(x) = \mathcal{E}_n\psi_n(x) ~,
\label{eq:dSE}
\end{equation}
whose lowest eigenvalue is, by construction, $\mathcal{E}_0 = 0$.
Also, the superpotential can be written in terms of $\psi_0(x)$ and $\eta(x)$ as following:
\begin{equation}
W(x) = -\frac{1}{\sqrt{2m_0}}\left( \eta(x)\frac{d}{dx}\ln\psi_0(x) + \frac{1}{2}\frac{d\eta(x)}{dx} \right) ~.
\end{equation}

We say the system is deformed shape invariant when $W(x)=W(x;\bm{a})$, where $\bm{a}$ denotes the set of parameters, satisfies
\begin{equation}
W(x;\bm{a})^2 + \frac{\eta(x)}{\sqrt{2m_0}}\frac{dW(x;\bm{a})}{dx}
= W(x;f(\bm{a}))^2 - \frac{\eta(x)}{\sqrt{2m_0}}\frac{dW(x;f(\bm{a}))}{dx} + \epsilon (\bm{a}) ~.
\label{eq:DSIcond}
\end{equation}
One notable feature of this class of exactly solvable problems is that, under the change of variable: 
\begin{equation}
x \mapsto z = z(x) \coloneqq \kappa \int^x\frac{d\bar{x}}{\eta(\bar{x})} 
\quad\text{\it i.e.}\quad
\frac{dz}{dx} = \frac{\kappa}{\eta(x)} ~,
\label{eq:x2z}
\end{equation}
the deformed shape-invariant Schr\"{o}dinger equation \eqref{eq:dSE} is mapped to a Schr\"{o}dinger equation with a conventional shape-invariant potential, say $U(z)$:
\begin{equation}
\left[ -\frac{1}{2m_0}\frac{d^2}{dz^2} + U(z)\right]\phi_n(z) = \varepsilon_n\phi_n(z) ~.
\end{equation}
Here, the potentials and the energy eigenvalues are related by
\begin{equation}
V(x) = \kappa^2U(z) ~,~~~
\mathcal{E}_{n} = \kappa^2\varepsilon_n ~.
\label{eq:PotEne_corresp}
\end{equation}
Moreover, during the transformation, the wave function transforms as
\begin{equation}
\psi_n(x) \mapsto \phi_n(z) \coloneqq \sqrt{\eta(x(z))}\,\psi_n(x(z)) ~.
\end{equation}

\subsection{Extension of SWKB quantization condition}
In our previous treatments on SWKB quantization condition, we have taken the unit of $\hbar = 2m = 1$ for simplicity.
Since, in this section, we are to discuss mass-dependency, we retain $m$ in the SWKB quantization condition:
\begin{equation}
\int_a^b \sqrt{2m\left[ \mathcal{E}_n - W(x)^2 \right]} \,dx = n\pi ~.
\label{eq:SWKB-m}
\end{equation}

One might guess that mere application of the SWKB condition to the deformed shape-invariant systems:
\begin{equation}
\int_a^b \sqrt{2m_0 [\mathcal{E}_n - W(x)^2]} \,dx \overset{?}{=} n\pi ~.
\end{equation}
would work, but this is not the case (See Example \ref{ex:dSWKB-dHO}).
By virtue of the relation between a deformed shape-invariant system and a conventional shape-invariant one \eqref{eq:x2z}, we extend the SWKB integral in the following manner:
\begin{equation}
\int_a^b \sqrt{2m_0 [\mathcal{E}_n - W(x)^2]} \,dx
\quad\to\quad
\int_a^b \sqrt{2m_0 [\mathcal{E}_n - W(x)^2]} \,\frac{dx}{\eta(x)} ~.
\end{equation}
Therefore, the extended version of the SWKB condition equation reads
\begin{equation}
\int_a^b \sqrt{2m_0 [\mathcal{E}_{{\rm d},n} - W(x)^2]} \,\frac{dx}{\eta(x)} = n\pi ~,~~~
n = 0,1,2,\ldots  ~.
\label{eq:extSWKB2}
\end{equation}
This extension can also be seen as the replacement of $m$ in Eq. \eqref{eq:SWKB-m} by $m_0/\eta(x)^2$.
These are straightforward from the interpretations of the deformed systems.
Note that this is an \textit{exact} condition equation for any deformed conventional shape-invariant potentials.

We have just justified the extension, so now we demonstrate the exactness.
The superpotential for a conventional shape-invariant potential $U(z)$ is expressed in terms of that for the corresponding deformed shape-invariant system in the following way:
\begin{equation}
w(z) = -\frac{1}{\kappa}W(x(z)) ~.
\label{eq:wW}
\end{equation}
Therefore, the extended SWKB integral becomes
\begin{equation}
\int_a^b \sqrt{2m_0 [\mathcal{E}_n - W(x)^2]} \,\frac{dx}{\eta(x)}
= \int_{\alpha}^{\beta} \sqrt{2m_0 [\varepsilon_n - w(z)^2]} \,dz 
= n\pi ~.
\end{equation}

\begin{example}[Deformed harmonic oscillator]
\label{ex:dSWKB-dHO}
We put $2m_0=1$ again for simplicity.
The simplest example of deformed shape-invariant system would be the deformed harmonic oscillator:
\begin{equation}
W(x) = \omega x ~,~~~
\eta(x) = 1 + \alpha x^2 ~,~~~
\mathcal{E}_n = 2n\omega + n^2\alpha ~.
\end{equation}
The ordinary SWKB integral \eqref{eq:SWKB-m} yields
\begin{equation}
\int_{-a}^{a}\sqrt{2n\omega + n^2\alpha - \omega^2x^2}\,dx 
= n\pi\left( 1 + \frac{n\alpha}{2\omega} \right) 
\neq n\pi ~.
\end{equation}
On the other hand, the extended SWKB integral \eqref{eq:extSWKB2} is
\begin{equation}
\int_{-a}^{a}\frac{\sqrt{2n\omega + n^2\alpha - \omega^2x^2}}{1+\alpha x^2}\,dx 
= 2\omega\int_0^a \frac{\sqrt{a^2-x^2}}{1+\alpha x^2} \,dx
= n\pi ~.
\end{equation}
\end{example}

\subsection{Another example: Semi-confined harmonic oscillator}
There is another example of one-dimensional quantum-mechanical systems with position-dependent effective mass, which is refered to as \textit{semi-confined harmonic oscillator}~\cite{Jafarov-2021,Jafarov:2022aa}.
This system is not deformed shape invariant, but the main part of the eigenfunctions are Laguerre polynomials.
Before concluding this Paper, we demonstrate that our extended SWKB condition \eqref{eq:extSWKB2} is also exactly applicable to this system.

\begin{example}[Semi-confined harmonic oscillator]
The superpotential, the position-dependent effective masses and the energy eigenvalues of this system are given by
\begin{equation}
W(x) = \begin{cases}
	\omega x\sqrt{\dfrac{x_0}{x+x_0}} & x>-x_0 \\
	-\infty & x\leqslant -x_0
\end{cases} ~,~~~ 
\eta(x) =  \begin{cases}
	\sqrt{\dfrac{x+x_0}{x_0}} & x>-x_0 \\
	\infty & x\leqslant -x_0
\end{cases} ~,~~~
\mathcal{E}_n = 2n\omega ~.
\end{equation}
Then, the extended SWKB integral reads
\begin{equation}
\int_{a_-}^{a_+} \sqrt{\frac{x_0}{x+x_0}\left( 2n\omega - \frac{x_0\omega^2x^2}{x+x_0} \right)} \,dx
= \omega x_0\int_{a'_-}^{a'_+} \sqrt{(y-a'_-)(a'_+-y)} \,\frac{dy}{y}
= n\pi ~,
\end{equation}
with
\begin{equation}
x+x_0 \equiv y ~,~~~
a'_{\pm} = \frac{n \pm \sqrt{n^2+2\omega x_0^2n}}{\omega x_0} - x_0 ~.
\end{equation}
\end{example}

\section{Conclusion}
\label{sec:Conclusion}
In this paper, first we have explored the interrelations among exactly solvable problems in one-dimensional quantum-mechanical systems in terms of both Schr\"{o}dinger equations and SWKB integrals. 
All the potentials belonging to the conventional shape-invariant potentials, where the eigenfunctions of the Hamiltonians are expressed in terms of the classical orthogonal polynomials, are closely connected with each other via the changes of variables, or the point canonical transformations. 
We have provided the comprehensive lists of the transformations. 
Note that while these relations among the Schr\"{o}dinger equations have been known for decades, this work is the first to demonstrate that similar correspondence also holds at the level of SWKB integrals.

Moreover, based on the interrelations, we have developed new quantization rules that are exact for Natanzon-type potentials. 
This class includes all the possible potentials whose eigenfunctions contain either the Laguerre or the Jacobi polynomial as the main part. 
Since most of the Natanzon-type potentials and hence the Schr\"{o}dinger equations are givenonly implicitly, our quantization conditions are also in implicit forms. 
The exactness of the condition equation is also confirmed through the explicit numerical calculations. 
These quantization rules can be regarded as the extension of the SWKB quantization condition, which is known to be exact only for the conventional shape-invariant potentials. 

We have also shown that the same idea of extending the SWKB quantization condition is also applicable to the quantum systems with position-dependent effective masses. 
For the systems whose eigenfunctions involve the classical orthogonal polynomials, we have also derived an exact quantization condition by suitably modifying the SWKB formula. 
Several explicit examples have been provided to illustrate the validity of the extension.

At the end of this paper, we conjecture what the SWKB quantization condition and its extensions we have presented throughout this paper imply. 
Based on the discussions above, we arrive at the following conjecture.
\begin{quote}\it
There exists $n \in \mathbb{Z}_{\geqslant 0}$ such that the SWKB quantization condition holds exactly if and only if the main part of the wavefunction $\phi_n$ with the same $n$, as well as the ground-state wavefunction $\phi_0$, are expressed in terms of the classical orthogonal polynomials with different orders.
\end{quote}
In short, the exactness of the SWKB quantization condition means the system is exactly solvable by the classical orthogonal polynomials. 
This statement is a conjecture in the sense that the mathematical proof is absent, but to the best of the author's knowledge, the conjecture is consistent with all known examples to date.


\bigskip
\section*{Acknowledgment}
The author would like to thank Nobuyuki Sawado and Ryu Sasaki for valuable discussions and their careful and useful comments.

\bibliographystyle{naturemag}
\bibliography{bibliography}

\begin{thebibliography}{10}
\expandafter\ifx\csname url\endcsname\relax
  \def\url#1{\texttt{#1}}\fi
\expandafter\ifx\csname urlprefix\endcsname\relax\def\urlprefix{URL }\fi
\providecommand{\bibinfo}[2]{#2}
\providecommand{\eprint}[2][]{\url{#2}}

\bibitem{alma9910116860608801}
\bibinfo{author}{Koekoek, R.}, \bibinfo{author}{Lesky, P.~A.} \&
  \bibinfo{author}{Swarttouw, R.~F.}
\newblock \emph{\bibinfo{title}{Hypergeometric orthogonal polynomials and their
  $q$-analogues}}.
\newblock Springer monographs in mathematics (\bibinfo{publisher}{Springer},
  \bibinfo{address}{Berlin}, \bibinfo{year}{2010}), \bibinfo{edition}{1st ed.}
  edn.

\bibitem{routh1884some}
\bibinfo{author}{Routh, E.~J.}
\newblock \bibinfo{title}{{On some properties of certain solutions of a
  differential equation of the second order}}.
\newblock \emph{\bibinfo{journal}{Proceedings of the London Mathematical
  Society}} \textbf{\bibinfo{volume}{1}}, \bibinfo{pages}{245--262}
  (\bibinfo{year}{1884}).

\bibitem{Bochner:1929aa}
\bibinfo{author}{Bochner, S.}
\newblock \bibinfo{title}{{\"{U}ber Sturm--Liouvillesche Polynomsysteme}}.
\newblock \emph{\bibinfo{journal}{Mathematische Zeitschrift}}
  \textbf{\bibinfo{volume}{29}}, \bibinfo{pages}{730--736}
  (\bibinfo{year}{1929}).
\newblock \urlprefix\url{https://doi.org/10.1007/BF01180560}.

\bibitem{Sasaki_2016}
\bibinfo{author}{Sasaki, R.} \& \bibinfo{author}{Znojil, M.}
\newblock \bibinfo{title}{{One-dimensional Schr\"{o}dinger equation with
  non-analytic potential $V(x) = -\exp (-|x|)$ and its exact Bessel-function
  solvability}}.
\newblock \emph{\bibinfo{journal}{Journal of Physics A: Mathematical and
  Theoretical}} \textbf{\bibinfo{volume}{49}}, \bibinfo{pages}{445303}
  (\bibinfo{year}{2016}).
\newblock \urlprefix\url{https://dx.doi.org/10.1088/1751-8113/49/44/445303}.

\bibitem{doi:10.1142/S0219887825400304}
\bibinfo{author}{Nasuda, Y.}
\newblock \bibinfo{title}{Bessel functions of purely imaginary order and an
  exactly solvable quantum-mechanical potential}.
\newblock \emph{\bibinfo{journal}{Int. J. Geom. Methods Mod. Phys.}}
  \textbf{\bibinfo{volume}{??}}, \bibinfo{pages}{2540030}
  (\bibinfo{year}{2025}).

\bibitem{krein1957continuous}
\bibinfo{author}{Krein, M.~G.}
\newblock \bibinfo{title}{{On a continuous analogue of a Christoffel formula
  from the theory of orthogonal polynomials}}.
\newblock In \emph{\bibinfo{booktitle}{Doklady Akademii Nauk}}, vol.
  \bibinfo{volume}{113}, \bibinfo{pages}{970--973}
  (\bibinfo{organization}{Russian Academy of Sciences}, \bibinfo{year}{1957}).

\bibitem{adler1994modification}
\bibinfo{author}{Adler, V.~{\'E}.}
\newblock \bibinfo{title}{{A modification of Crum's method}}.
\newblock \emph{\bibinfo{journal}{Theoretical and Mathematical Physics}}
  \textbf{\bibinfo{volume}{101}}, \bibinfo{pages}{1381--1386}
  (\bibinfo{year}{1994}).

\bibitem{sasaki2010exceptional}
\bibinfo{author}{Sasaki, R.}, \bibinfo{author}{Tsujimoto, S.} \&
  \bibinfo{author}{Zhedanov, A.}
\newblock \bibinfo{title}{{Exceptional Laguerre and Jacobi polynomials and the
  corresponding potentials through Darboux--Crum transformations}}.
\newblock \emph{\bibinfo{journal}{Journal of Physics A: Mathematical and
  Theoretical}} \textbf{\bibinfo{volume}{43}}, \bibinfo{pages}{315204}
  (\bibinfo{year}{2010}).

\bibitem{ODAKE2011164}
\bibinfo{author}{Odake, S.} \& \bibinfo{author}{Sasaki, R.}
\newblock \bibinfo{title}{{Exactly solvable quantum mechanics and infinite
  families of multi-indexed orthogonal polynomials}}.
\newblock \emph{\bibinfo{journal}{Physics Letters B}}
  \textbf{\bibinfo{volume}{702}}, \bibinfo{pages}{164--170}
  (\bibinfo{year}{2011}).
\newblock
  \urlprefix\url{https://www.sciencedirect.com/science/article/pii/S0370269311007416}.

\bibitem{GOMEZULLATE2009352}
\bibinfo{author}{G{\'o}mez-Ullate, D.}, \bibinfo{author}{Kamran, N.} \&
  \bibinfo{author}{Milson, R.}
\newblock \bibinfo{title}{{An extended class of orthogonal polynomials defined
  by a Sturm--Liouville problem}}.
\newblock \emph{\bibinfo{journal}{Journal of Mathematical Analysis and
  Applications}} \textbf{\bibinfo{volume}{359}}, \bibinfo{pages}{352--367}
  (\bibinfo{year}{2009}).
\newblock
  \urlprefix\url{https://www.sciencedirect.com/science/article/pii/S0022247X09004569}.

\bibitem{GOMEZULLATE2010987}
\bibinfo{author}{G{\'o}mez-Ullate, D.}, \bibinfo{author}{Kamran, N.} \&
  \bibinfo{author}{Milson, R.}
\newblock \bibinfo{title}{{An extension of Bochner's problem: Exceptional
  invariant subspaces}}.
\newblock \emph{\bibinfo{journal}{Journal of Approximation Theory}}
  \textbf{\bibinfo{volume}{162}}, \bibinfo{pages}{987--1006}
  (\bibinfo{year}{2010}).
\newblock
  \urlprefix\url{https://www.sciencedirect.com/science/article/pii/S0021904509001853}.

\bibitem{quesne2008exceptional}
\bibinfo{author}{Quesne, C.}
\newblock \bibinfo{title}{{Exceptional orthogonal polynomials, exactly solvable
  potentials and supersymmetry}}.
\newblock \emph{\bibinfo{journal}{Journal of Physics A: Mathematical and
  Theoretical}} \textbf{\bibinfo{volume}{41}}, \bibinfo{pages}{392001}
  (\bibinfo{year}{2008}).

\bibitem{ODAKE2009414}
\bibinfo{author}{Odake, S.} \& \bibinfo{author}{Sasaki, R.}
\newblock \bibinfo{title}{{Infinitely many shape invariant potentials and new
  orthogonal polynomials}}.
\newblock \emph{\bibinfo{journal}{Physics Letters B}}
  \textbf{\bibinfo{volume}{679}}, \bibinfo{pages}{414--417}
  (\bibinfo{year}{2009}).
\newblock
  \urlprefix\url{https://www.sciencedirect.com/science/article/pii/S0370269309009186}.

\bibitem{ODAKE2010173}
\bibinfo{author}{Odake, S.} \& \bibinfo{author}{Sasaki, R.}
\newblock \bibinfo{title}{{Another set of infinitely many exceptional
  ($X_{\ell}$) Laguerre polynomials}}.
\newblock \emph{\bibinfo{journal}{Physics Letters B}}
  \textbf{\bibinfo{volume}{684}}, \bibinfo{pages}{173--176}
  (\bibinfo{year}{2010}).
\newblock
  \urlprefix\url{https://www.sciencedirect.com/science/article/pii/S0370269310000158}.

\bibitem{Odake_2011}
\bibinfo{author}{Odake, S.} \& \bibinfo{author}{Sasaki, R.}
\newblock \bibinfo{title}{Discrete quantum mechanics}.
\newblock \emph{\bibinfo{journal}{Journal of Physics A: Mathematical and
  Theoretical}} \textbf{\bibinfo{volume}{44}}, \bibinfo{pages}{353001}
  (\bibinfo{year}{2011}).
\newblock \urlprefix\url{https://dx.doi.org/10.1088/1751-8113/44/35/353001}.

\bibitem{natanzon1971study}
\bibinfo{author}{Natanzon, G.~A.}
\newblock \bibinfo{title}{{Study of the one-dimensional Schroedinger equation
  generated from the hypergeometric equation}}.
\newblock \emph{\bibinfo{journal}{Vestnik Leningradskogo Universiteta}}
  \bibinfo{pages}{22--28} (\bibinfo{year}{1971}).

\bibitem{natanzon1979general}
\bibinfo{author}{Natanzon, G.~A.}
\newblock \bibinfo{title}{{General properties of potentials for which the
  Schr\"{o}dinger equation can be solved by means of hypergeometric
  functions}}.
\newblock \emph{\bibinfo{journal}{Theoretical and Mathematical Physics}}
  \textbf{\bibinfo{volume}{38}}, \bibinfo{pages}{146--153}
  (\bibinfo{year}{1979}).

\bibitem{ginocchio1984class}
\bibinfo{author}{Ginocchio, J.~N.}
\newblock \bibinfo{title}{{A class of exactly solvable potentials. I.
  One-dimensional Schr\"{o}dinger equation}}.
\newblock \emph{\bibinfo{journal}{Annals of Physics}}
  \textbf{\bibinfo{volume}{152}}, \bibinfo{pages}{203--219}
  (\bibinfo{year}{1984}).

\bibitem{cooper1987relationship}
\bibinfo{author}{Cooper, F.}, \bibinfo{author}{Ginocchio, J.~N.} \&
  \bibinfo{author}{Khare, A.}
\newblock \bibinfo{title}{{Relationship between supersymmetry and solvable
  potentials}}.
\newblock \emph{\bibinfo{journal}{Physical Review D}}
  \textbf{\bibinfo{volume}{36}}, \bibinfo{pages}{2458} (\bibinfo{year}{1987}).

\bibitem{quesne-SIGMA2007}
\bibinfo{author}{Quesne, C.}
\newblock \bibinfo{title}{{Point Canonical Transformation versus Deformed Shape
  Invariance for Position-Dependent Mass Schr{\"o}dinger Equations}}.
\newblock \emph{\bibinfo{journal}{SIGMA. Symmetry, Integrability and Geometry:
  Methods and Applications}} \textbf{\bibinfo{volume}{5}}, \bibinfo{pages}{046}
  (\bibinfo{year}{2009}).
\newblock \urlprefix\url{https://www.emis.de/journals/SIGMA/2009/046/}.

\bibitem{PhysRev.152.683}
\bibinfo{author}{BenDaniel, D.~J.} \& \bibinfo{author}{Duke, C.~B.}
\newblock \bibinfo{title}{{Space-Charge Effects on Electron Tunneling}}.
\newblock \emph{\bibinfo{journal}{Phys. Rev.}} \textbf{\bibinfo{volume}{152}},
  \bibinfo{pages}{683--692} (\bibinfo{year}{1966}).
\newblock \urlprefix\url{https://link.aps.org/doi/10.1103/PhysRev.152.683}.

\bibitem{PhysRevB.27.7547}
\bibinfo{author}{von Roos, O.}
\newblock \bibinfo{title}{{Position-dependent effective masses in semiconductor
  theory}}.
\newblock \emph{\bibinfo{journal}{Phys. Rev. B}} \textbf{\bibinfo{volume}{27}},
  \bibinfo{pages}{7547--7552} (\bibinfo{year}{1983}).
\newblock \urlprefix\url{https://link.aps.org/doi/10.1103/PhysRevB.27.7547}.

\bibitem{Bagchi_2005}
\bibinfo{author}{Bagchi, B.}, \bibinfo{author}{Banerjee, A.},
  \bibinfo{author}{Quesne, C.} \& \bibinfo{author}{Tkachuk, V.~M.}
\newblock \bibinfo{title}{{Deformed shape invariance and exactly solvable
  Hamiltonians with position-dependent effective mass}}.
\newblock \emph{\bibinfo{journal}{Journal of Physics A: Mathematical and
  General}} \textbf{\bibinfo{volume}{38}}, \bibinfo{pages}{2929}
  (\bibinfo{year}{2005}).
\newblock \urlprefix\url{https://dx.doi.org/10.1088/0305-4470/38/13/008}.

\bibitem{Jafarov-2021}
\bibinfo{author}{Jafarov, E.~I.} \& \bibinfo{author}{Van~der Jeugt, J.}
\newblock \bibinfo{title}{{Exact solution of the semiconfined harmonic
  oscillator model with a position-dependent effective mass}}.
\newblock \emph{\bibinfo{journal}{The European Physical Journal Plus}}
  \textbf{\bibinfo{volume}{136}}, \bibinfo{pages}{758} (\bibinfo{year}{2021}).
\newblock \urlprefix\url{https://doi.org/10.1140/epjp/s13360-021-01742-z}.

\bibitem{Jafarov:2022aa}
\bibinfo{author}{Jafarov, E.~I.} \& \bibinfo{author}{Van~der Jeugt, J.}
\newblock \bibinfo{title}{{Exact solution of the semiconfined harmonic
  oscillator model with a position-dependent effective mass in an external
  homogeneous field}}.
\newblock \emph{\bibinfo{journal}{Pramana}} \textbf{\bibinfo{volume}{96}},
  \bibinfo{pages}{35} (\bibinfo{year}{2022}).
\newblock \urlprefix\url{https://doi.org/10.1007/s12043-021-02279-7}.

\bibitem{witten1981dynamical}
\bibinfo{author}{Witten, E.}
\newblock \bibinfo{title}{{Dynamical breaking of supersymmetry}}.
\newblock \emph{\bibinfo{journal}{Nuclear Physics B}}
  \textbf{\bibinfo{volume}{188}}, \bibinfo{pages}{513--554}
  (\bibinfo{year}{1981}).

\bibitem{witten1982constraints}
\bibinfo{author}{Witten, E.}
\newblock \bibinfo{title}{{Constraints on supersymmetry breaking}}.
\newblock \emph{\bibinfo{journal}{Nuclear Physics B}}
  \textbf{\bibinfo{volume}{202}}, \bibinfo{pages}{253--316}
  (\bibinfo{year}{1982}).

\bibitem{cooper1995supersymmetry}
\bibinfo{author}{Cooper, F.}, \bibinfo{author}{Khare, A.} \&
  \bibinfo{author}{Sukhatme, U.}
\newblock \bibinfo{title}{{Supersymmetry and quantum mechanics}}.
\newblock \emph{\bibinfo{journal}{Physics Reports}}
  \textbf{\bibinfo{volume}{251}}, \bibinfo{pages}{267--385}
  (\bibinfo{year}{1995}).

\bibitem{gendenshtein1983derivation}
\bibinfo{author}{Gendenshtein, L.~E.}
\newblock \bibinfo{title}{{Derivation of Exact Spectra of the Schr\"{o}dinger
  equation by means of supersymmetry}}.
\newblock \emph{\bibinfo{journal}{Jetp Lett}} \textbf{\bibinfo{volume}{38}},
  \bibinfo{pages}{356--359} (\bibinfo{year}{1983}).

\bibitem{gangopadhyaya2017supersymmetric}
\bibinfo{author}{Gangopadhyaya, A.}, \bibinfo{author}{Mallow, J.~V.} \&
  \bibinfo{author}{Rasinariu, C.}
\newblock \emph{\bibinfo{title}{{Supersymmetric quantum mechanics: An
  introduction}}} (\bibinfo{publisher}{World Scientific Publishing Company},
  \bibinfo{year}{2017}).

\bibitem{comtet1985exactness}
\bibinfo{author}{Comtet, A.}, \bibinfo{author}{Bandrauk, A.} \&
  \bibinfo{author}{Campbell, D.~K.}
\newblock \bibinfo{title}{{Exactness of semiclassical bound state energies for
  supersymmetric quantum mechanics}}.
\newblock \emph{\bibinfo{journal}{Physics Letters B}}
  \textbf{\bibinfo{volume}{150}}, \bibinfo{pages}{159--162}
  (\bibinfo{year}{1985}).

\bibitem{doi:10.1142/S0218271825300010}
\bibinfo{author}{Gangopadhyaya, A.}, \bibinfo{author}{Bougie, J.} \&
  \bibinfo{author}{Rasinariu, C.}
\newblock \bibinfo{title}{Recent advances regarding the exactness of
  semiclassical methods inspired by supersymmetric quantum mechanics}.
\newblock \emph{\bibinfo{journal}{International Journal of Modern Physics D}}
  \textbf{\bibinfo{volume}{34}}, \bibinfo{pages}{2530001}
  (\bibinfo{year}{2025}).
\newblock \urlprefix\url{https://doi.org/10.1142/S0218271825300010}.
\newblock \eprint{https://doi.org/10.1142/S0218271825300010}.

\bibitem{dutt1986exactness}
\bibinfo{author}{Dutt, R.}, \bibinfo{author}{Khare, A.} \&
  \bibinfo{author}{Sukhatme, U.~P.}
\newblock \bibinfo{title}{{Exactness of supersymmetric WKB spectra for
  shape-invariant potentials}}.
\newblock \emph{\bibinfo{journal}{Physics Letters B}}
  \textbf{\bibinfo{volume}{181}}, \bibinfo{pages}{295--298}
  (\bibinfo{year}{1986}).

\bibitem{khare1989shape}
\bibinfo{author}{Khare, A.} \& \bibinfo{author}{Varshni, Y.}
\newblock \bibinfo{title}{{Is shape invariance also necessary for lowest order
  supersymmetric WKB to be exact?}}
\newblock \emph{\bibinfo{journal}{Physics Letters A}}
  \textbf{\bibinfo{volume}{142}}, \bibinfo{pages}{1--4} (\bibinfo{year}{1989}).

\bibitem{delaney1990susy}
\bibinfo{author}{DeLaney, D.} \& \bibinfo{author}{Nieto, M.~M.}
\newblock \bibinfo{title}{{SUSY-WKB is neither exact nor never worse than WKB
  for all solvable potentials}}.
\newblock \emph{\bibinfo{journal}{Physics Letters B}}
  \textbf{\bibinfo{volume}{247}}, \bibinfo{pages}{301--308}
  (\bibinfo{year}{1990}).

\bibitem{bougie2018supersymmetric}
\bibinfo{author}{Bougie, J.}, \bibinfo{author}{Gangopadhyaya, A.} \&
  \bibinfo{author}{Rasinariu, C.}
\newblock \bibinfo{title}{{The supersymmetric WKB formalism is not exact for
  all additive shape invariant potentials}}.
\newblock \emph{\bibinfo{journal}{Journal of Physics A: Mathematical and
  Theoretical}} \textbf{\bibinfo{volume}{51}}, \bibinfo{pages}{375202}
  (\bibinfo{year}{2018}).

\bibitem{Nasuda:2020aqf}
\bibinfo{author}{Nasuda, Y.} \& \bibinfo{author}{Sawado, N.}
\newblock \bibinfo{title}{{Numerical study of the SWKB condition of novel
  classes of exactly solvable systems}}.
\newblock \emph{\bibinfo{journal}{Mod. Phys. Lett. A}}
  \textbf{\bibinfo{volume}{36}}, \bibinfo{pages}{2150025}
  (\bibinfo{year}{2021}).
\newblock \eprint{2004.04927}.

\bibitem{10.1007/978-981-19-4751-3_29}
\bibinfo{author}{Nasuda, Y.}
\newblock \bibinfo{title}{{Several Exactly Solvable Quantum Mechanical Systems
  and the SWKB Quantization Condition}}.
\newblock In \bibinfo{editor}{Dobrev, V.} (ed.) \emph{\bibinfo{booktitle}{Lie
  Theory and Its Applications in Physics}}, \bibinfo{pages}{339--349}
  (\bibinfo{publisher}{Springer Nature Singapore},
  \bibinfo{address}{Singapore}, \bibinfo{year}{2022}).

\bibitem{NASUDA2023116087}
\bibinfo{author}{Nasuda, Y.} \& \bibinfo{author}{Sawado, N.}
\newblock \bibinfo{title}{{SWKB quantization condition for conditionally
  exactly solvable systems and the residual corrections}}.
\newblock \emph{\bibinfo{journal}{Nuclear Physics B}}
  \textbf{\bibinfo{volume}{987}}, \bibinfo{pages}{116087}
  (\bibinfo{year}{2023}).
\newblock
  \urlprefix\url{https://www.sciencedirect.com/science/article/pii/S0550321323000160}.

\bibitem{bhalla1996exactness}
\bibinfo{author}{Bhalla, R.}, \bibinfo{author}{Kapoor, A.} \&
  \bibinfo{author}{Panigrahi, P.}
\newblock \bibinfo{title}{{Exactness of the supersymmetric WKB approximation
  scheme}}.
\newblock \emph{\bibinfo{journal}{Physical Review A}}
  \textbf{\bibinfo{volume}{54}}, \bibinfo{pages}{951} (\bibinfo{year}{1996}).

\bibitem{bhalla1997quantum}
\bibinfo{author}{Bhalla, R.}, \bibinfo{author}{Kapoor, A.} \&
  \bibinfo{author}{Panigrahi, P.}
\newblock \bibinfo{title}{{Quantum Hamilton--Jacobi formalism and the bound
  state spectra}}.
\newblock \emph{\bibinfo{journal}{American Journal of Physics}}
  \textbf{\bibinfo{volume}{65}}, \bibinfo{pages}{1187--1194}
  (\bibinfo{year}{1997}).

\bibitem{leacock1983hamilton_action}
\bibinfo{author}{Leacock, R.~A.} \& \bibinfo{author}{Padgett, M.~J.}
\newblock \bibinfo{title}{{Hamilton--Jacobi theory and the quantum action
  variable}}.
\newblock \emph{\bibinfo{journal}{Physical Review Letters}}
  \textbf{\bibinfo{volume}{50}}, \bibinfo{pages}{3} (\bibinfo{year}{1983}).

\bibitem{leacock1983hamilton}
\bibinfo{author}{Leacock, R.~A.} \& \bibinfo{author}{Padgett, M.~J.}
\newblock \bibinfo{title}{{Hamilton-Jacobi/action-angle quantum mechanics}}.
\newblock \emph{\bibinfo{journal}{Physical Review D}}
  \textbf{\bibinfo{volume}{28}}, \bibinfo{pages}{2491} (\bibinfo{year}{1983}).

\bibitem{gozzi1986nodal}
\bibinfo{author}{Gozzi, E.}
\newblock \bibinfo{title}{{Nodal structure of supersymmetric wave functions}}.
\newblock \emph{\bibinfo{journal}{Physical Review D}}
  \textbf{\bibinfo{volume}{33}}, \bibinfo{pages}{3665} (\bibinfo{year}{1986}).

\bibitem{cooper2001supersymmetry}
\bibinfo{author}{Cooper, F.}, \bibinfo{author}{Khare, A.} \&
  \bibinfo{author}{Sukhatme, U.~P.}
\newblock \emph{\bibinfo{title}{{Supersymmetry in quantum mechanics}}}
  (\bibinfo{publisher}{World Scientific}, \bibinfo{year}{2001}).

\bibitem{sym12111853}
\bibinfo{author}{Quesne, C.}
\newblock \bibinfo{title}{{Deformed Shape Invariant Superpotentials in Quantum
  Mechanics and Expansions in Powers of $\hbar$}}.
\newblock \emph{\bibinfo{journal}{Symmetry}} \textbf{\bibinfo{volume}{12}}
  (\bibinfo{year}{2020}).
\newblock \urlprefix\url{https://www.mdpi.com/2073-8994/12/11/1853}.

\end{thebibliography}

\end{document}